\documentclass[11pt]{article}
\usepackage{epsfig}
\usepackage{amssymb,amsmath}
\usepackage{epigraph}
\usepackage{xcolor}

\usepackage[lined,boxed,commentsnumbered]{algorithm2e}

\usepackage{graphicx}				
\usepackage{amssymb}
\usepackage{multirow}

\usepackage{siunitx,booktabs}
\usepackage{newtxtext,newtxmath}

\def\boldeta{\text{\boldmath$\eta$}}

\def\bold_eta{\text{\boldmath$\eta$}}

\newcommand{\beq}{\begin{equation}}
\newcommand{\eeq}{\end{equation}}
\newcommand{\bea}{\begin{eqnarray}}
\newcommand{\eea}{\end{eqnarray}}

\begin{document}



\begin{center}

{\LARGE
Non-Linear and Meta-Stable Dynamics in Financial Markets: \\
\vskip0.5cm
Evidence from High Frequency  Crypto Currency Market Makers 

}

\vskip1.0cm
{\Large Igor Halperin\footnote{Fidelity Investments. 
 Opinions expressed here are author's own, and do not represent views of his employer. A standard disclaimer applies. E-mail for communications on the paper: ighalp@gmail.com}  
} \\
\vskip0.5cm

\today

\vskip1.0cm
{\Large Abstract:\\}
\end{center}
\parbox[t]{\textwidth}{
This work builds upon the long-standing conjecture that linear diffusion models are inadequate for complex market dynamics. Specifically, it provides experimental validation for the author's prior arguments that realistic market dynamics are governed by higher-order (cubic and higher) non-linearities in the drift. 
As the diffusion drift is given by the negative gradient of a potential function, this means that a non-linear drift translates into a non-quadratic potential. These arguments were based both on general theoretical grounds as well as a structured approach to modeling the price dynamics which incorporates money flows and their impact on market prices. Here, we find direct confirmation of this view by analyzing high-frequency crypto currency data at different time scales ranging from minutes to months. We find that markets can be characterized by either a single-well or a double-well potential, depending on the time period and sampling frequency, where a double-well potential may signal market uncertainty or stress.
}
 
 \newcounter{helpfootnote}
\setcounter{helpfootnote}{\thefootnote} 
\renewcommand{\thefootnote}{\fnsymbol{footnote}}
\setcounter{footnote}{0}
\footnotetext{
I would like to thank  Galin Georgiev for providing me the data and many valuable discussions. I also thank Andrey Itkin for collaboration on research related to topics raised in this paper, and helpful comments on the draft. 
}     

 \renewcommand{\thefootnote}{\arabic{footnote}}
\setcounter{footnote}{\thehelpfootnote} 

\newpage
 
\section{Introduction}

The challenge of accurately modeling the complex dynamics of financial asset prices has long been a central theme in quantitative finance. While linear diffusion models, such as the Geometric Brownian Motion (GBM), provide a foundational and often tractable framework employed in particular in the celebrated Black-Scholes option pricing model \cite{BS}, they are insufficient to capture the full richness of market behavior. While the inadequacy of simple linear models such as the GBM is widely acknowledged, the overwhelming majority of efforts by both academics and practitioners has focused on modifying the noise term, leading to sophisticated stochastic and local volatility models, jump-diffusions, Lev{\'y} processes etc., but has rarely touched alternative models with non-linear drifts.  

The idea of non-linearity of price dynamics is not new. One of the first known to the author proposals of this sort was made by Jan Dash \cite{Dash_book}, who suggested that linear price dynamics is just too simple to match the complex dynamics of financial markets, and proposed a non-linear Reggeon diffusion model.\footnote{As noted in \cite{Dash_book}, the author sadly had to discontinue his research when asked by his manager if he was ready to bet his bonus on the success of his non-linear diffusion model. Industrial research is a risky business.}
 
The theoretical underpinning of the present work follows this same line of inquiry. It stems from a series of papers by the author and co-authors \cite{HD_QED, Inverted_World, NES, MANES, Marketron, Marketron_options} that propose a paradigm where the most significant non-linearities in market dynamics originate from the drift term itself. This framework, motivated by a first-principles analysis of money flows and their price impact, posits that the drift is a non-linear function of a state variable $ x $ such as the current asset price or its log-price. In the language of physics, this is equivalent to stating that the system evolves within a non-quadratic potential $U(x)$. The drift, or "force," acting on the system is given by the negative gradient of this potential, $\mu(x) = -\partial U(x)/\partial x$. Detailing the shape and parameters of the potential  $ U(x) $, by either deriving them from a theory or estimating them from data, is therefore critical for understanding properties of market dynamics.

While these models were motivated by first-principles arguments connecting money flows to market impact, their core predictions, i.e. the existence and specific forms of non-linear drifts and non-quadratic potentials, require a direct experimental validation. Previous experiments conducted with either daily equity returns \cite{HD_QED, Marketron} or equity index and single-stock options \cite{NES, MANES, Marketron_options} suggested that for daily equity data, market-implied potentials usually have a single-well shape, but sometimes may turn into a double-well shape, especially during periods of enhanced market volatility or/and instability. Potentials having a double-well shape require a cubic power dependence of the drift on the state variable. This suggests that the price dynamics are generally driven by non-quadratic potentials, while deviations from the behavior corresponding to a quadratic potential (and hence a linear drift) are typically seen (become essential) only during periods of market instability and/or enhanced volatility.

The main limitation of daily equity return data is its sparsity. To collect enough data points to estimate our non-linear models, we need to include long time periods measured in months or even years. For such time scales the assumption that the drift of price diffusion depends only on the price itself might be too heroic: many other factors (macro-economic,  political, etc.) impact the behavior of price dynamics at such longer scales. On the other hand, as the principles leading to the prediction of non-linear drifts and non-quadratic potentials are general and not specific to the daily-traded equity markets, one might be interested in exploring non-linearities in financial data collected at higher frequencies.     

This line of inquiry has recently gained traction. For example, Wand et al. \cite{Wand2024} used a maximum-likelihood framework on intraday stock market data specifically to test for the presence of a quartic potential (corresponding to a cubic drift) as proposed in \cite{HD_QED}. Interestingly, their model selection criteria found strong statistical support for a cubic potential (quadratic drift) in their datasets, but not for the higher-order quartic potential. This highlights the importance of using a model-free approach, as the market-implied potential may vary across different markets and time regimes, and demonstrates the need to investigate the distinct types of market dynamics at different time scales and in different market regimes.

To this end, we turn to the high-frequency world of cryptocurrency markets, which serve as an ideal real-world laboratory for such study. These markets, particularly the Automated Market Makers (AMMs) in Decentralized Finance (DeFi), offer an unparalleled source of rich, granular, and publicly available data on transactions and liquidity.

Our analysis focuses on data from Uniswap v3 pools on the Arbitrum network, a leading platform for decentralized exchange \cite{Uniswapv3}. Uniswap v3 is characterized by its "concentrated liquidity" mechanism, where liquidity providers can specify narrow price ranges for their capital. This design creates a highly state-dependent liquidity profile, making it a fertile ground for observing the non-linear dynamics we seek to uncover \cite{Georgiev2024}.

Instead of assuming a specific parametric model, we employ a non-parametric approach using the Kramers-Moyal expansion, well known in statistical physics, see e.g. \cite{Gardiner_book}. This technique allows us to directly estimate the drift and diffusion functions from the empirical time-series data of the asset's log-price. From the estimated physical drift rate, we then compute the underlying potential $U(x)$ by numerical integration. By performing this analysis across multiple time scales from minutes to hours and days, we construct a multi-dimensional picture of the market's underlying non-linear dynamics.

The results presented in this work offer a direct, model-free confirmation of our central hypothesis. We find clear evidence of highly non-linear drift functions whose non-linearity gradually weakens with a decreased sampling frequency. Consequently, the derived potential functions are distinctly non-quadratic, exhibiting well-defined minima that act as equilibrium or quasi-equilibrium points for the price. We show how the shape and depth of these potential wells evolve over time and across different market periods, sometimes revealing a single-well structure indicative of a stable regime, and at other times hinting at a more complex, multi-well landscape that may signal market uncertainty or stress.
This work thus bridges the gap between the theoretical framework of non-linear market dynamics and its direct, empirical observation in one of today's most dynamic financial ecosystems.

Our paper is organized as follows. In Sect.~\ref{sect_data}, we describe our datasets and methodology. Results of our estimations are presented in Sect.~\ref{sect_results}. The next Sect.~\ref{sect_discussion} presents discussion of our results, and the final Sect.~\ref{sect_conclusion} concludes. 

\section{Data and Experimental Setup}
\label{sect_data}

\subsection{Data Source and Pre-processing}
The primary data for this study consists of high-frequency transaction data from liquid Uniswap v3 pools on the Arbitrum chain. We analyze three datasets. The first dataset contains the history of the USDC-WETH log-price on Arbitrum from 01-01-2024 to 12-31-2024, with about 300k transaction records. The second dataset is for the same USDC-WETH pair for the period 01-01-2025 to 06-30-2025. The third dataset is for the WBTC-WETH pair for the period 01-01-2024 to 12-31-2024.

These pools represent two fundamentally different types of markets. The \textbf{USDC-WETH} pool is the crypto-equivalent of a major fiat currency pair, such as EUR/USD. It prices a highly volatile and central ecosystem asset (WETH, or Wrapped Ethereum) against a stablecoin (USDC) that is pegged to the US Dollar. The dynamics of this pool are therefore anchored to a stable frame of reference, reflecting the value of the Ethereum ecosystem in dollar terms.

In contrast, the \textbf{WBTC-WETH} pool is a "crypto-cross" pair, representing the relative valuation of the two largest and most significant cryptocurrencies: Bitcoin (in its wrapped, ERC-20 compatible form) and Ether. In this market, \textit{both} assets are highly volatile, and there is no stable anchor. Its dynamics are driven by the complex interplay of factors affecting the two dominant crypto-assets relative to each other.

Note that the two markets are weakly correlated with each other. While both {USDC-WETH and WBTC-WETH are 
technically spread, the first one is against a stable coin, so it factually represents the price
of (wrapped) Ethereum (WETH). In contrast,  WBTC-WETH is the wrapped Bitcoin-Ethereum spread.\footnote{ 
For analogies with the equity ETF markets,
the USDC-WETH pool is similar to SPY, while 
the WBTC-WETH would be similar to the spread between SPY and IWM ETFs. I thank Galin Georgiev for 
educating me on Uniswap v3 and Arbitrum network.}
By analyzing these two distinct market types—one anchored to a stable value and one representing a relative valuation of two volatile assets—we can explore how their underlying potential landscapes differ and evolve under various conditions.

Each row in our raw datasets represents the log-price of the asset pair, averaged over 15 consecutive transactions to provide a slightly smoothed time-series that reduces microscopic noise due to low-informative low-volume transactions. The datasets include a high-precision timestamp for each observation, with a typical time-step between consecutive records being around 100 seconds. To mitigate the influence of extreme data points that can distort non-parametric estimation, we trim the data by removing the top 0.5\% and bottom 0.5\% of observations based on their log-price values. 

\subsection{Methodology}
Our analysis is based on the Langevin equation, 
a standard representation for stochastic processes in physics, see e.g. \cite{Gardiner_book}. We model the evolution of the log-price, denoted by $x(t)$, using the Langevin equation
\begin{equation}
\label{Langevin}
    \dot{x}(t) = \mu(x) + \sigma(x)\xi_t
\end{equation}
Here, $\dot{x}(t)$ is the time derivative of the log-price. The dynamics are governed by two state-dependent functions: the drift $\mu(x)$, which represents a deterministic force, and the volatility (or diffusion coefficient) $\sigma(x)$, which modulates the strength of the random fluctuations. The term $\xi_t$ is a Gaussian white noise process with zero mean and delta-correlation, $\langle \xi_t \xi_{t'} \rangle = \delta(t-t')$.

The Langevin equation (\ref{Langevin}) is equivalent to the It{\^ o}'s stochastic differential equation (SDE) commonly used in finance. By formally multiplying both sides of (\ref{Langevin}), we can write it as follows
\begin{equation}
\label{Itos_diffusion}
    dx = \mu(x)dt + \sigma(x)dW_t
\end{equation}
where $dW_t = \xi_t dt$ is the increment of a standard Wiener process. This equivalence allows us to bridge the physical Langevin picture with standard financial modeling based on It{\^o}'s diffusion (\ref{Itos_diffusion}).

While in general the drift function $\mu(x)$ might depend on external factors (i.e. itself be stochastic), in this work we focus on the settings when it can be reasonably approximated by a deterministic function of the current state variable $ x = x_t $.\footnote{Essentially, this assumption is equivalent to the assumption that the time intervals at which we analyze the price dynamics,
all other stochastic factors are either averaged, or assumed to be slowly varying and hence approximately constant during observation periods.}  
The deterministic drift function $\mu(x)$  is directly related to the underlying self-interaction potential $U(x)$ of the system, representing the force that pulls the system towards a global equilibrium or a local minimum:
\begin{equation}
\label{potential}
    \mu(x) = -\frac{\partial U(x)}{\partial x}
\end{equation}
A linear drift corresponds to a simple quadratic (harmonic) potential\footnote{Note that the standard Ornstein-Uhlenbeck (OU) process has a linear drift and a corresponding quadratic potential.}, while the non-linear drift we seek to uncover corresponds to a more complex, non-quadratic potential.

We use the `kramersmoyal` Python package \cite{KM_package} to non-parametrically estimate the Kramers-Moyal coefficients, $K_1(x)$ and $K_2(x)$ (see e.g. \cite{Gardiner_book}), from the time-series data. These are related to the physical drift and diffusion rates via the average time step, $\Delta t$, within an analyzed window:
\begin{equation}
    \mu(x) = \frac{K_1(x)}{\Delta t}, \quad \frac{\sigma^{2}(x)}{2} = \frac{K_2(x)}{2 \Delta t}
\end{equation}
These physical rates are then annualized for financial interpretation. The potential $U(x)$ is computed using Eq.(\ref{potential}) by numerically integrating the estimated physical drift rate $\mu(x)$. To investigate the dynamics across different time horizons, we perform a multi-lag analysis by subsampling the time-series at different `lag` values, which correspond to observing the system over real-world time intervals from minutes to hours.

\section{Results}
\label{sect_results}

We conducted our analysis over several periods in 2024 and 2025 to investigate the stability and evolution of the market's dynamic properties. In this section, we first present results separately for all our datasets, and then present analysis of our findings.

\subsection{The USDC-WETH 2024 pool}

The results for three representative periods are consolidated in Figure \ref{fig:multi_period_analysis_3USDC_WETH_2024}. Each row in the figure corresponds to a different time window of length 2 months, while each column shows the estimated annualized drift $ \mu(x) $, annualized volatility $ \sigma(x) $, and the derived potential $ U(x) $, respectively. Within each plot, the different colored lines represent the analysis performed at different time scales (lags), from 10 minutes to 6 hours. 

The most striking feature of these graphs is a strong non-linearity of the drift (the left column) which translates into a double-well potential $ U(x) $ for two out of three time periods analyzed in these plots (see the right column in Fig.~\ref{fig:multi_period_analysis_3USDC_WETH_2024}).
Very similar potential profiles with a pronounced double-well structure
are observed for windows of 3 or 4 months (not reported here to save space.)

What happens if we take shorter windows, such as one month or shorter?
To illustrate the dependence of results on the lookback window, in Fig.~\ref{fig:multi_period_analysis_3USDC_WETH_2024_short}, we show results onbtained with one-month windows periods. One can see that the double well potential observed for longer time windows is gone and replaced by a (non-harmonic) single-well potential. We suggest that this is due to data insufficiency in the tails for such small windows. For this reason, the subsequent analysis in this section will focus on window sizes of 2 month only.

\begin{figure}[h!]
    \centering
    \includegraphics[width=\textwidth]{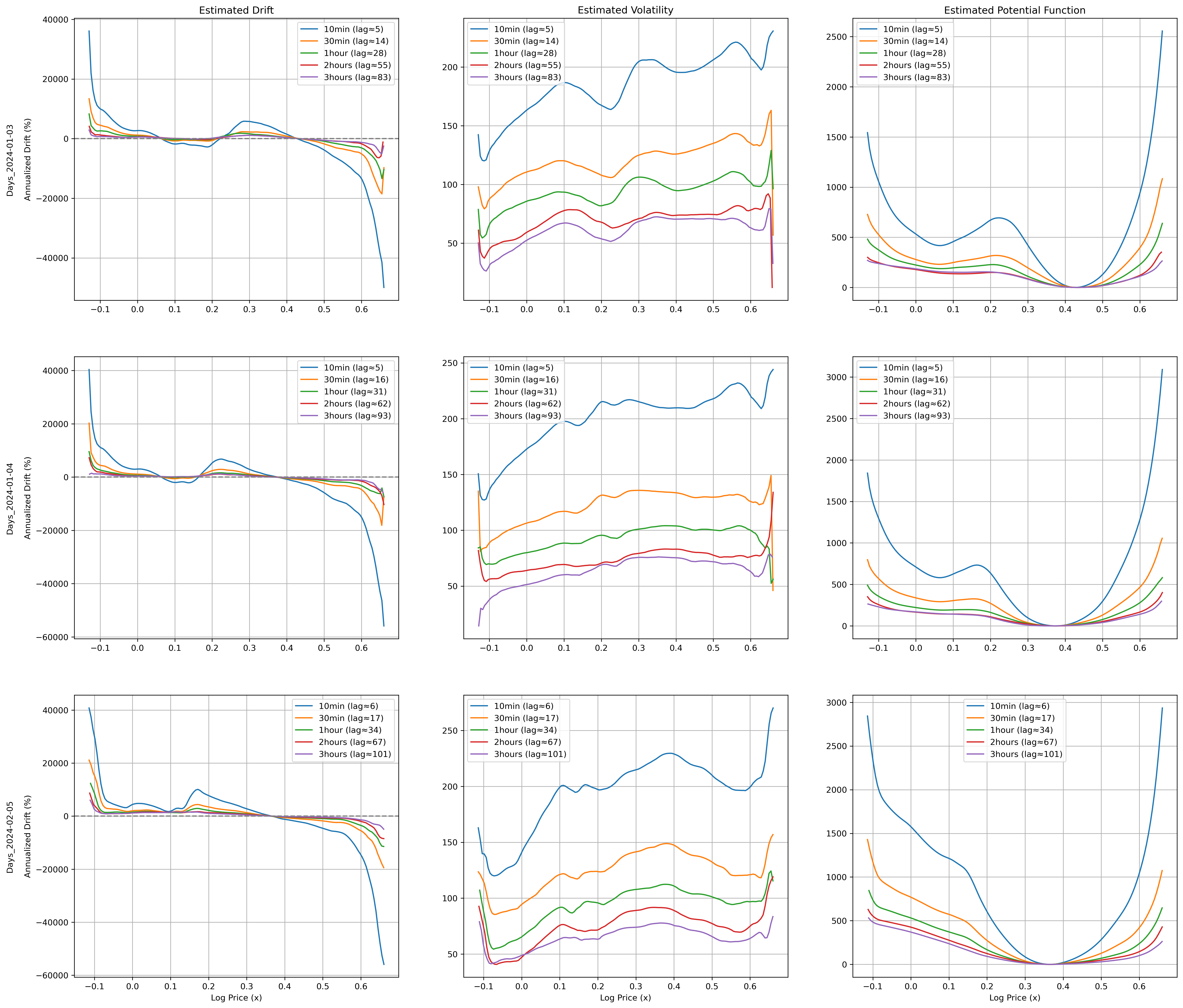} 
    \caption{
        Comparison of estimated drift,  volatility, and potential $ U(x) $ for the USDC-WETH log-price on Arbitrum in 2024.
  Each row represents a distinct monthly time period, and each curve within a plot correspond to a different sampling frequency from 10 min to 3 hours. 
        The potential $U(x)$ is derived from the physical drift rate $\mu(x)$ and normalized to have a minimum of zero.
    }
    \label{fig:multi_period_analysis_3USDC_WETH_2024}
\end{figure}

\begin{figure}[h!]
    \centering
    \includegraphics[width=\textwidth]{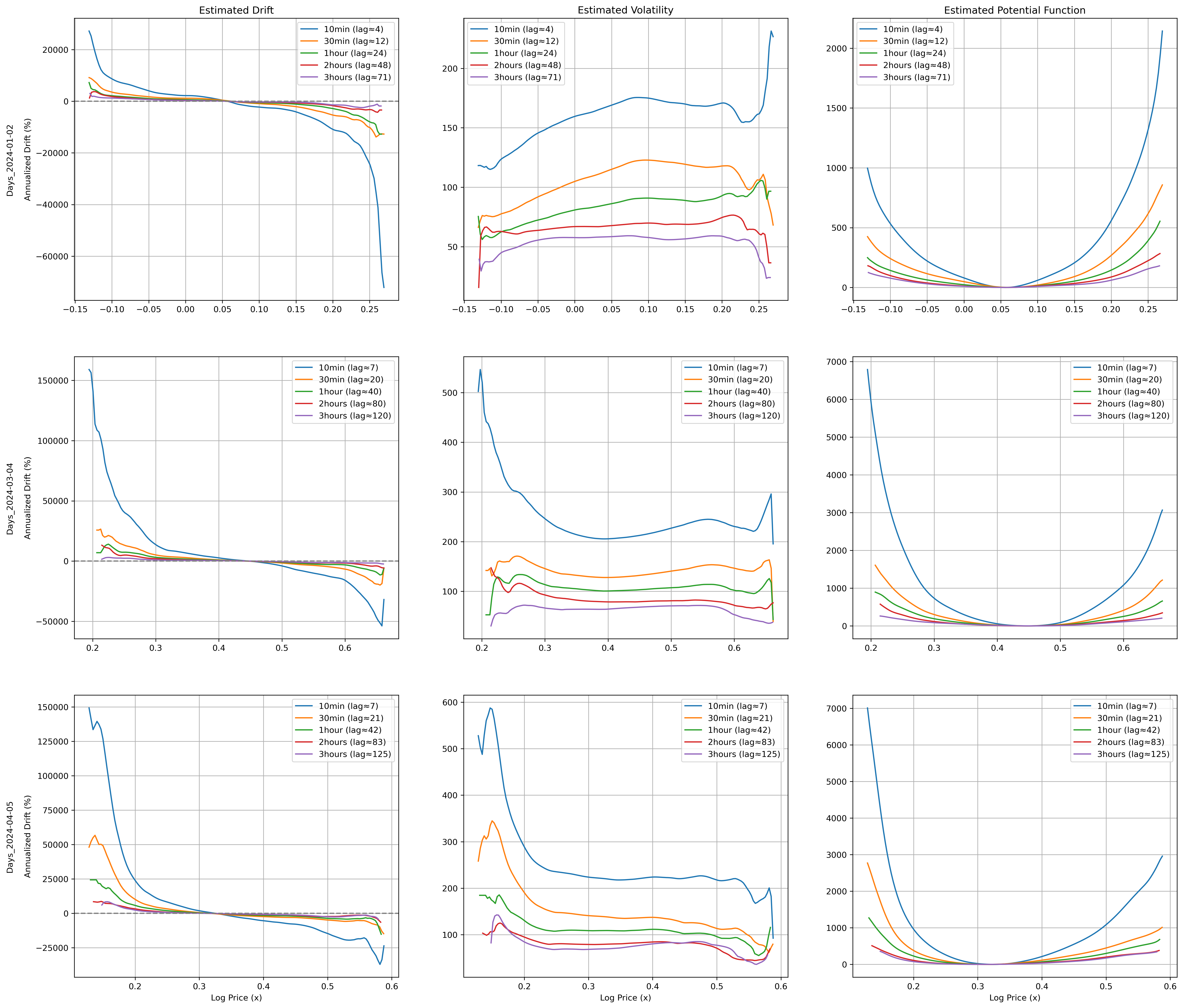} 
    \caption{
        Comparison of estimated drift,  volatility, and potential $ U(x) $ for the USDC-WETH log-price on Arbitrum in 2024, for shorter windows.
 Each row represents a distinct monthly time period, and each curve within a plot correspond to a different sampling frequency from 10 min to 3 hours. 
        The potential $U(x)$ is derived from the physical drift rate $\mu(x)$ and normalized to have a minimum of zero.
    }
    \label{fig:multi_period_analysis_3USDC_WETH_2024_short}
\end{figure}

\subsection{The UBTC-WETH 2024 pool}

The results for three representative periods are consolidated in Figure \ref{fig:multi_period_analysis_3UBTC_WETH_2024}.
Again, each row in the figure corresponds to a different time window of length 2 months, while each column shows the estimated annualized drift, annualized volatility, and the derived self-interaction potential $ U(x) $, respectively. Within each plot, the different colored lines represent the analysis performed at different time scales (lags), from 10 minutes to 6 hours. 

Unlike the  USDC-WETH 2024 pool dataset, in this case,
produced drift and especially potential shapes have a different pattern of producing single-well (but still non-quadratic) potentials. 

\begin{figure}[h!]
    \centering
    \includegraphics[width=\textwidth]{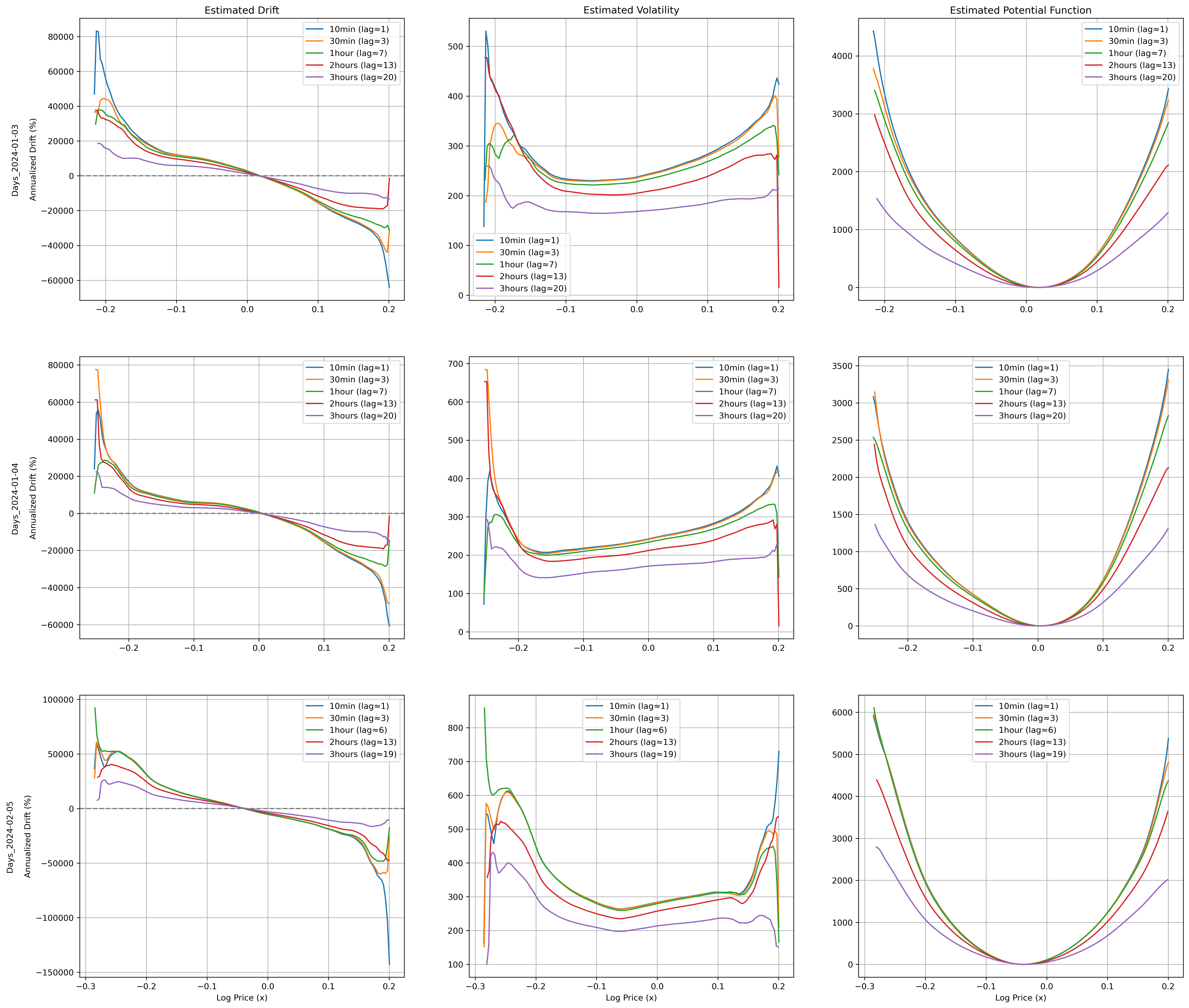} 
    \caption{
        Comparison of estimate drift, volatility, and potential $ U(x) $ for the UBTC-WETH log-price on Arbitrum in 2024.
         Each row represents a distinct  2 months time period, and each curve within a plot correspond to a different sampling frequency from 10 min to 3 hours.
        The potential $U(x)$ is derived from the physical drift rate $\mu(x)$ and normalized to have a minimum of zero.
    }
    \label{fig:multi_period_analysis_3UBTC_WETH_2024}
\end{figure}


\subsection{The USDC-WETH 2025 pool} 

To compare the market behavior during different periods, we additionally analyzed the behavior 
of the  USDC-WETH pool during the first six months of 2025.
The results for three representative periods are presented in the same format as above in Figure \ref{fig:multi_period_analysis_3UBTC_WETH_2025}.

Comparing with the analyis of the same pool in 2024, the  USDC-WETH 2025 pool dataset suggests yet different patterns of strongly non-quadratic potentials 
$ U(x) $. As can be seen in the right column, some of the produced potentials,
especially for shorter sampling frequencies, have a small local minimum to the right of a stable global minimum, while in other periods or for other sampling frequencies, the potential is of a single-well type. However, in all cases for this dataset, the potential has a very wide basin, suggesting a strong non-linearity.  

\begin{figure}[h!]
    \centering
    \includegraphics[width=\textwidth]{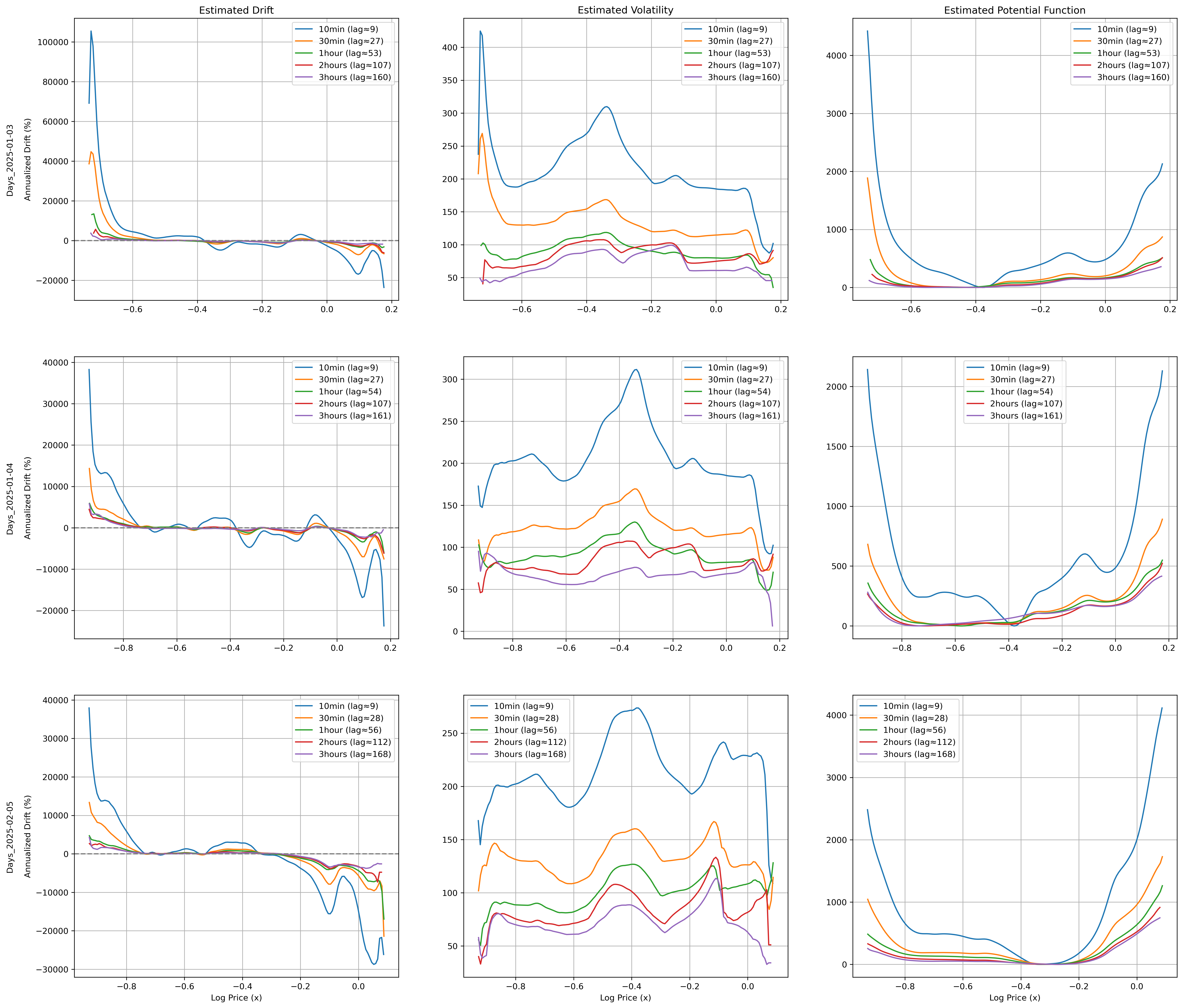} 
    \caption{
        Comparison of estimated drift, volatility, and potential $ U(x) $ for the UBTC-WETH log-price on Arbitrum in 2025.
        Each row represents a distinct  2 months time period, and each curve within a plot correspond to a different sampling frequency from 10 min to 3 hours.
     The potential $U(x)$ is derived from the physical drift rate $\mu(x)$ and normalized to have a minimum of zero.
    }
    \label{fig:multi_period_analysis_3UBTC_WETH_2025}
\end{figure}

\section{Discussion}
\label{sect_discussion}


We thus analyzed several two-months (and multi-days) periods in 2024 and 2025 across both the USDC-WETH and WBTC-WETH pools to investigate the stability and evolution of the market's dynamic properties. 
Here we present a summary analysis for results consolidated across Figures 1-4.

\subsubsection{Confirmation of Non-Quadratic and Meta-Stable Potentials}

Our findings across all datasets and time periods consistently confirm the core hypothesis of the non-linear market models \cite{HD_QED, NES, MANES, Marketron}: that {\bf the dynamics are driven by generally non-quadratic potentials}.
Depending on the market regime (period) and sampling frequency, the self-interaction potentials $ U(x) $ can be of a single-well or a double-well type (with occasional multi-well shapes suggested for some datasets).
For low-frequency sampling (above hourly time steps), 
potentials $ U(x) $ typically have a single global minimum, and sometimes small local minima to its right.
Even though they typically demonstrate deviations from 
a harmonic (quadratic) potential, unless they have a very wide basin as in Fig.~\ref{fig:multi_period_analysis_3UBTC_WETH_2025}, their dynamics may still be reasonably well approximated by a quadratic (harmonic oscillator) potential. 

One the other hand, at shorter sampling frequencies (under an hour), we observe a richer variety of behaviors. The derived potential functions are dynamic, producing a slow evolution between two profiles:
\begin{itemize}
    \item A {\bf single well potential}, indicative of a strong mean-reverting force stabilizing the price around a clear equilibrium point (a resilient market state).
    \item A {\bf double-well potential} or multi-well potential, indicative of the system having a meta-stable state where two competing price regimes or fixed points exist. This may signal increased market uncertainty or stress, as the price is not firmly anchored to one dominant equilibrium, see also Sect.~\ref{sect_instantons} below for further discussion.
\end{itemize}
For shorter time windows (one month and shorter), the pattern for the resulting potential flips, and now we get single well potential for all time windows, as illustrated in Fig.~\ref{fig:multi_period_analysis_3USDC_WETH_2024_short}. 
As was mentioned above, such visibly different behavior 
may be explained by data insufficiency in the tails of distributions for such small windows. 


\subsubsection{Any Evidence for a Stochastic Drift?}

As mentioned above, one possible explanation of different potential shapes obtained for longer ($ \geq 2 $ months) and short ($ \leq 1 $ month) observation windows is due to data insufficiency in the tails for short windows. 
Note that this argument, as well as the whole analysis performed so far, assumes that the drift $ \mu(x) $ is a deterministic function of log-price $ x $, and not a function of some additional hidden stochastic factors.

Here we explore an alternative scenario that may exist: that the drift function $\mu(x)$ contains {\bf hidden stochastic components} $ \boldeta_{t} $. Specifically, the drift might be composed of:
\begin{itemize}
\item {\it Short-lived stochastic terms} ($\eta_{s}$): Driven by fast-moving microstructure or external news events.
\item {\it Slow-varying stochastic terms} ($\eta_{l}$): Driven by longer-term fundamental, macroeconomic, or political factors.
\end{itemize}
When we use a {\it short observation window} (e.g., one month), we capture the rapid fluctuations caused by $\eta_{s}$. If these high-frequency factors are strong enough, they are expected to introduce substantial noise into the drift estimate, potentially obscuring the underlying long-term structure and making the potential appear unstable or flat across a short time span. This is {\it not} what we see in the data for short windows, suggesting that if any fast hidden stochastic factors are present in $ \mu(x) $, they settle at their equilibrium distribution (i.e. self-average) at the scale of days or even hours.

When we use a {\it long observation window} (e.g., two months), the $\eta_{s}$ terms are already averaged out. The resulting drift estimate is cleaner and reflects the structural properties driven by the slow-varying signals $\eta_{l}$ terms. In its turn, their evolution leads to a slowly changing potential $ U(x) $ that can go back and forth between a double-well and a single-well shape when constructed for high sampling frequencies (under an hour).

To summarize, our results appear consistent with the hypothesis that if any hidden stochastic terms are present in the drift $ \mu(x) $, it is only slowly-varying types of them which are of practical interest, as any fast-changing stochastic factors are effectively averaged out and replaced by their mean in drift and potential estimations for windows exceeding 5 days.


\subsection{Implications for model builders and physics aficionados}
\label{sect_instantons}

Our experiments demonstrated that non-linear effects in the drift are very significant for the intraday price dynamics, especially at intraday-appropriate time steps (under one hour). Depending on the market regime and possibly on slowly varying hidden stochastic terms in the drift, the potential implied by these non-linear drift can be of either single-well or double-well type. Here we briefly outline implications of each one of such scenario.

For a single-well potential, the dynamics in a vicinity of a stable global minimum of the potential can still be approximated by a quadratic potential.\footnote{Again, as long as the potential does not have a very wide basin as in 
Fig.~\ref{fig:multi_period_analysis_3UBTC_WETH_2025}.}
Non-linearities are only important if quantities that we care about are very sensitive to the tail behavior of the log-price distribution.
      
On the other hand, for non-linear drifts that produce {\bf double-well} potentials $ U(x) $, implications for the dynamics are more profound. The presence of local minima of potentials gives rise to {\bf metastable dynamics}. Such dynamics are most easily explained using the physics language. If we associate the log-price $ x $ with the position of a 'particle' (referred to as the marketron in \cite{Marketron}), the particle can get into and out of such local minimum as a result of thermal fluctuations.

The key observation is that such a state of the system residing around its local minimum is a {\it metastable} state - its lifetime depends on the height of a potential barrier separating the global and local minima, see 
Fig.~\ref{fig:multi_period_analysis_3USDC_WETH_2024}.
For a sufficiently high barrier, such metastable state can live long, making escape from such state a rare event.
While such escape may happen more or less rarely depending on the potential shape, the transition itself occurs very fast, almost instantaneously. For this reason, this type of transitions between local and global minima are known in physics as {\bf instantons}. While the relevance of 
the physics of double wells and instantons for modeling market dynamics were highlighted in 
\cite{HD_QED, NES, MANES, Marketron} based on theoretical grounds, in the present paper we find a direct evidence of relevance of instantons (as 'features' of double-well potentials) for intraday price dynamics.
If the potential barrier is not too high, instanton transitions can occur at higher rates, contributing to enhanced price volatility. On the other hand, if the double well potential and instantons are appropriate for the dynamics but overlooked by model developers, they would be forced to associate a higher price volatility with an increased volatility of the drving Brownian motion.

\section{Conclusion}
\label{sect_conclusion}

This work presented a direct, non-parametric analysis of high-frequency cryptocurrency price data, providing strong experimental evidence for the theories of non-linear market dynamics previously proposed by the author. By applying the Kramers-Moyal expansion to extensive datasets from Uniswap v3, we have moved beyond parametric assumptions and directly measured the drift and diffusion functions governing price evolution. The estimated drift was further used to find the potentials describing the market dynamics under different conditions. 

Our findings confirm that the drift of the log-price often shows large deviations from linearity, especially at short time scales. Consequently, the potential derived from such drift is distinctly non-quadratic, while its shape and depth are dependent on the observation time scale and the market regime. For time steps of an hour and shorter, we find that the dynamics may slowly drift between single-well and double-well potentials. For lower sampling frequencies, the potential is almost always of a single-well type, though it remains distinctly non-harmonic. This dependence of the potential's topology on the sampling frequency is a powerful manifestation of the complex, multi-scale, and non-linear dynamics that govern high-frequency crypto markets.

\end{document}